\documentclass[12pt]{article}
\pdfoutput=1

\usepackage{setspace}
\usepackage{amsmath}
\usepackage{graphicx}

\usepackage{amsfonts}
\usepackage[german,english]{babel}
\usepackage{bm,pstricks}                          % bold math

\newcommand{\kB}{k_{\rm B}}

\begin{document}

\title{Thermal expansion in nanoresonators}

\author{ Agust\'in Mancardo Viotti$^{1}$, Alejandro G. Monastra $^{1,2,*}$, \\ Mariano F. Moreno$^{1}$, M. Florencia Carusela$^{1,2}$}

\maketitle

$^1$ Instituto de Ciencias, Universidad Nacional de Gral. Sarmiento, J. M. Guti\'errez 1150, (1613) Los Polvorines, Argentina\\
$^{2}$ Consejo Nacional de Investigaciones Cient\'ificas y T\'ecnicas, Godoy Cruz 2290, (1425) Buenos Aires, Argentina

%\date{}

$^{*}$Email: amonast@ungs.edu.ar

\begin{abstract} 

Inspired by some recent experiments and numerical works related to nanoresonators, we perform classical molecular dynamics simulations to investigate the thermal expansion and the ability of the device to act as a strain sensor assisted by thermally-induced vibrations.  The proposed model consists in a chain of atoms interacting anharmonically with both ends clamped to thermal reservoirs. We analyze the thermal expansion and resonant frequency shifts as a function of temperature and the applied strain.  For the transversal modes the shift is approximately linear with strain. We also present analytical results from canonical calculations in the harmonic approximation showing that thermal expansion is uniform along the device. This prediction also works when the system operates in a nonlinear oscillation regime at moderate and high temperatures.
\end{abstract}

%\begin{keyword}
%transport processes/heat transfer, thermal expansion, low dimensional systems
%\end{keyword}

\maketitle

%%%%%%%%%%%%%%%%%%%%%%%%%%%%%%%%%%%%%%%%%%%%%%%%%%%%%%%%%%%%%%%%%%%%%%%%%
\section{Introduction}
%%%%%%%%%%%%%%%%%%%%%%%%%%%%%%%%%%%%%%%%%%%%%%%%%%%%%%%%%%%%%%%%%%%%%%%%%

Increasing interest in resonant motion of nanosystems is due to its promising capability to act as sensors.
Theoretical and experimental studies enabled advances not only in the characterization but also in the production of nanoscale resonators \cite{nems5, sensor3,sensor5}.

The functionality as sensors is based on their low masses, low force constants,  large resonant frequencies and low damping, i.e. high mechanical quality factors \cite{sensor1,sensor2, nems1, nems2, nems3, nems4}. Nanoscale resonators  are typically carbon-based structures, capable of weighing single bacteria \cite{sensor1}, detecting single spins in magnetic resonance systems \cite{sensor2}, and even probing quantum mechanics in macroscopic systems \cite{sensor3}. In particular, carbon nanotubes and graphene resonators \cite{sensor7reson,sensor9} are nearly ideal building material due to their perfect atomic structure with low density and high Young’s modulus.

Another interesting feature of nanoresonators is their strongly nonlinear behavior \cite{sensor3, sensor7, sensor8}. This produces  nonlinear vibrational modes, localized and non-localized modes, enabling an energy transfer between them \cite{nonl1,nonl2}. These consequences directly affect their thermal properties such as thermal expansion, intrinsic thermal vibrations and conductivity.  In addition to nonlinearities and depending on the dimensionality of the device, the role played by longitudinal, transversal and flexural modes and the coupling between them will be critical when studying thermal properties. For example, it was shown that the graphene's superior thermal conductivity and the behaviour of  the coefficient of thermal expansion are mainly due to the interplay between its three acoustic phonon modes, the fundamental role of flexural modes on thermal fluctuations and their particular vibration morphology \cite{modes1}.

On the other hand, these nanomechanical carbon-based devices are usually suspended and thus subject to tensile stress, which leads to high mechanical stability and high mechanical quality factors. The applied mechanical stress also affects the vibrational modes frequencies, and consequently their thermal response.  This has led to the development of an elastic strain engineering in order to improve the performance of  the transport and sensor properties of the resonators. This technique is a low-cost way of continuously tuning the phonon and electronic modes and thus the desired material properties.  Just to mention an example, in a recent experimental work it was proposed a versatile “local-self-calibration” and nondestructive method to monitor the applied strains in semiconductor micro or nanostructures, where local strains can be measured through analyzing the relative position of Raman peaks \cite{raman}.

Harmonic and anharmonic models \cite{dhar,ren2} have been proposed to theoretically study the underlying physical mechanisms involved in the energy transfer. For example for carbon-nanotubes or nanowires devices, the models are in general structures built from one dimensional arrays of atoms that can vibrate only longitudinally. However, in order to get a more reliable picture of the energy transfer phenomena, one should include the longitudinal modes as well as transversal and flexural ones \cite{nuestro}.

We propose a model for nanoresonator which can act as a strain sensor while it undergoes a thermal expansion. 
In this work we model a nanowire as a chain of  atoms with a $\alpha - \beta$ Fermi-Pasta-Ulam interaction potential.
The Fermi-Pasta Ulam (FPU) model and its variants provide an ideal test-bed for addressing fundamental
issues in statistical mechanics such as the validity of macroscopic laws in low dimensional
systems \cite{DDN,das}, when strong nonlinearities have to be consider. Thermal properties, as thermal conductivity, have been extensively investigated in one dimensional chains of atoms oscillating in one direction using a FPU model , e.g. demonstrating a breakdown of the normal-diffusional Fourier’s law dynamics \cite{mai}. As we are interested in the role of different kinds of modes, we generalize the $\alpha - \beta$ FPU model, to the case of  particles that are allowed to oscillate in any direction enabling also transversal modes.

%%%%%%%%%%%%%%%%%%%%%%%%%%%%%%%%%%%%%%%%%%%%%%%%%%%%%%%%%%%%%%%%%%%%%%%%%
\section{System model}
%%%%%%%%%%%%%%%%%%%%%%%%%%%%%%%%%%%%%%%%%%%%%%%%%%%%%%%%%%%%%%%%%%%%%%%%%

\begin{figure}[ht]
\begin{center}
\includegraphics[scale=0.3]{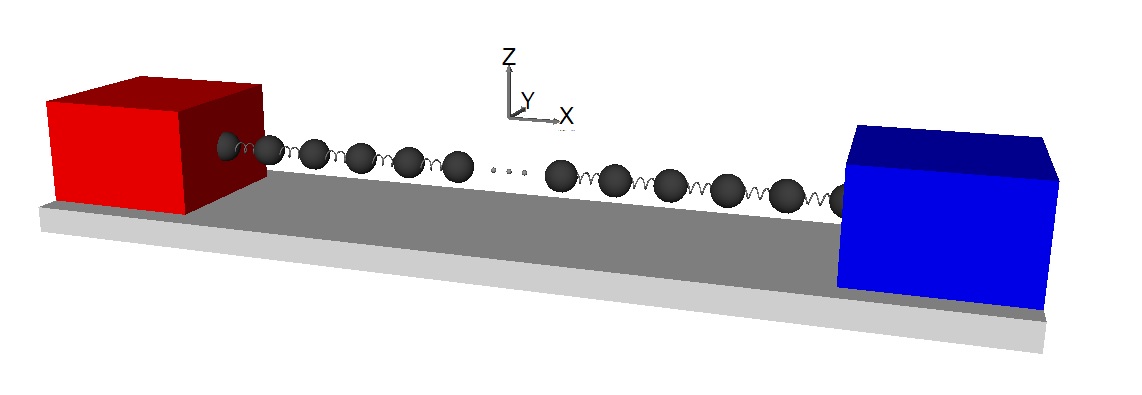}
\end{center}
\caption{Schematic of the system. Particles on each extreme are coupled to Langevin thermal baths.}
\label{figModel}
\end{figure}

We consider a system of $N$ particles in a linear arrange, identified by an index $1 \leq i \leq N$, at positions $\mathbf{R}_{i}$, which in principle can have different masses $m_i$. The particles interact to nearest neighbors by an $\alpha$-$\beta$ Fermi-Pasta-Ulam potential that only depends on the relative distance $d_i = | \mathbf{R}_{i+1} - \mathbf{R}_{i} |$

\begin{equation} \label{PotFPU}
v(d_i) = \frac{1}{2} k (d_i - l_0)^2 + \frac{1}{3} \alpha (d_i - l_0)^3 + \frac{1}{4} \beta (d_i - l_0)^4  \ .
\end{equation}
where $l_0$ is the equilibrium distance. The particles on the left $(i = 1)$ and right $(i = N)$ borders also interact by the same potential with two substrates that can be thought as a left $(i = 0)$ and right $(i = N + 1)$ fixed particles, so we are considering a nanowire with fixed boundary conditions. Therefore, there are $(N +1)$ interactions or bonds that contribute to the total potential

\begin{equation} \label{TotalPot}
V (\{ \mathbf{R}_{i} \} ) = \sum_{i = 0}^{N} v (  d_i ) \ ,
\end{equation}
that depends on the positions $ \mathbf{R}_{i}$. We allow for the particles to move in the three dimensions (see figure \ref{figModel}). A natural equilibrium position of the particles is a linear array along the $x$-axe with lattice constant $a = l_{0}$. However, if the distance between the fixed particles is bigger than $(N + 1) l_{0}$, there will be an uniaxial tension along the system that is parametrized by a change in the lattice constant $a > l_{0}$, or the strain $\varepsilon = (a-l_0)/l_0$. We are specially interested on the effects of tension on the thermal properties of the system because it is an external parameter that can be easily controlled. The equilibrium positions of the particles are $\mathbf{R}_{0 i} = ( i a , 0, 0)$. We characterize the motion of the particles by the displacement with respect to their equilibrium position $\mathbf{r}_i = \mathbf{R}_i - \mathbf{R}_{0 i} = (x_i, y_i, z_i)$. Moreover, the left and right particles $i=1$ and $N$ are coupled to two thermal reservoirs respectively. We consider a Langevin interaction by a viscous term proportional to velocity, and decorrelated random forces acting on the particles in contact with the reservoirs. The equation of motion for each particle is

\begin{equation} \label{Newton}
m_{i} \frac{\text{d}^2 \mathbf{r}_{i} }{\text{d} t^2}  = - \frac{\partial V }{\partial \mathbf{r}_{i} } - \gamma_{i} \frac{\text{d} \mathbf{r}_{i} }{\text{d} t} + \mathbf{f}_{i} (t)
\end{equation}
where $\gamma_{i} = \gamma \neq 0$ for $i =1$ or $N$, and zero otherwise. The random forces have the correlations

\begin{equation} \label{ForceCorrel}
\langle f_{ i, \mu} (t) f_{ j, \nu} (t') \rangle = 2 \gamma \ \kB T_{i} \ \delta_{i, j} \ \delta_{\mu, \nu} \ \delta (t - t')
\end{equation}
where $T_{i}$ is $T_{\text{L}}$ and  $T_{\text{R}}$ for $i =1$ and $N$, the temperatures of the left and right reservoirs, respectively, or zero otherwise. The indices $(\mu, \nu)$ run over the $(x,y,z)$ directions of motion. 

We focus on the thermal expansion of the nanowire, which we can compute from the average distance $d_i$, and on the resonances of transversal and longitudinal displacements of the atoms. The thermal expansion and resonances will strongly depend on temperature and tension.

%%%%%%%%%%%%%%%%%%%%%%%%%%%%%%%%%%%%%%%%%%%%%%%%%%%%%%%%%%%%%%%%%%%%%%%%%
\subsection{Expansion of the potential}
%%%%%%%%%%%%%%%%%%%%%%%%%%%%%%%%%%%%%%%%%%%%%%%%%%%%%%%%%%%%%%%%%%%%%%%%%

Considering the general case of particles moving in three directions and with tension, the nonlinear potential can be expanded around the new equilibrium positions for small displacements. The distance between two neigboring atoms is 

\begin{equation}
d_i = \sqrt{( a + \Delta_x )^2 + \Delta^2_{\perp} } \ ,
\end{equation}
where $\Delta_x = x_{i+1}-x_i$ and $\Delta^2_{\perp} = (y_{i+1}-y_i)^2  + (z_{i+1}-z_i)^2 $ are their relative longitudinal and transversal displacements. For both displacements being much smaller than the lattice constant $a$, we can expand it up to fourth order obtaining

\begin{equation} \label{Distance}
d_i = a + \Delta_x + \frac{1}{2 a} \Delta^2_{\perp} - \frac{1}{2 a^2}  \Delta_x \Delta^2_{\perp}  +  \frac{1}{8 a^3}  ( 4 \Delta^2_x \Delta^2_{\perp} -  \Delta^4_{\perp} ) + {\cal O} ( a^{-4} ) \ .
\end{equation}
Inserting this expansion into the potential (\ref{PotFPU}) we obtain

\begin{equation} \label{PotFPUExpanded}
v(d_i) = v_0 + F_0 \Delta_x + \frac{1}{2} k_{\text{eff}} \Delta^2_x + \frac{1}{2} k_{\perp} \Delta^2_{\perp} + \frac{1}{3} \alpha_{\text{eff}} \Delta^3_x + c_3  \Delta_x \Delta^2_{\perp} +  \frac{1}{4} \beta  \Delta^4_x +  \frac{1}{4} \beta_{\perp} \Delta^4_{\perp}  + c_4 \Delta^2_x \Delta^2_{\perp} \ ,
\end{equation}
with

\begin{eqnarray} \label{EffConstants}
v_0 &=& v(a) =  \frac{1}{2} k (a - l_0)^2 + \frac{1}{3} \alpha (a - l_0)^3 + \frac{1}{4} \beta (a - l_0)^4 \nonumber \\
F_0 &=&  k (a - l_0) + \alpha (a - l_0)^2 + \beta (a - l_0)^3 \nonumber \\ 
k_{\text{eff}} &=& k + 2 \alpha (a - l_0) + 3 \beta (a - l_0)^2 \nonumber \\ 
\alpha_{\text{eff}} &=& \alpha + 3 \beta (a - l_0) \nonumber \\
k_{\perp} &=& \frac{F_0}{a} \nonumber \\ 
\beta_{\perp} &=& \frac{c_3}{a} \nonumber \\  
c_3 &=& \frac{1}{2} k \frac{l_0}{a^2} + \frac{1}{2} \alpha \left( 1 - \frac{l^2_0}{a^2} \right) + \frac{1}{4} \beta  \left( 4 a -6 l_0 + 2 \frac{l^3_0}{a^2} \right) \nonumber \\ 
c_4 &=&  - \frac{1}{2} k \frac{l_0}{a^3} + \frac{1}{2} \alpha \frac{l^2_0}{a^3} + \frac{1}{2} \beta  \left( 1 - \frac{l^3_0}{a^3} \right)
\end{eqnarray}
$v_0$ is an energy constant with no effect on the dynamics. $F_0$ is the force in the longitudinal direction in between two neighboring atoms, i.e. the tension as a function of the lattice constant $a$, which indeed is nonlinear. $k_{\text{eff}}$  is an effective value of the force constant in the longitudinal direction. Depending on the values and sign of $\alpha$ it can increase or decrease with strain. $\alpha_{\text{eff}}$ replaces the $\alpha$ constant for the term proportional to the cube of the displacement in the longitudinal direction. $k_{\perp}$ represents a force constant, proportional to the tension, which is the leading term for the transversal direction. In case of no tension, this term vanishes, and the leading term in the transversal direction is quartic in the displacement and proportional to  $\beta_{\perp}$. $c_3$ and $c_4$ are two constants which couple the longitudinal and transversal coordinates at third and fourth order in the displacements, respectively. These terms are the responsible for the mixture and coupling of modes, reducing the transport of energy along the nanowire.

Summing these terms for every bond, the force terms proportional to $ \Delta_x$ cancel each other, obtaining up to second order

\begin{equation} \label{TotalPotApprox}
V (\{ \mathbf{r}_{i} \} ) = (N+1) v_0 +  \frac{1}{2} k_{\text{eff}} \sum_{i = 0}^{N} ( x_{i+1}-x_i)^2 +  \frac{1}{2} k_{\perp} \sum_{i = 0}^{N} [ (y_{i+1}-y_i)^2  + (z_{i+1}-z_i)^2 ] + \ldots \ .
\end{equation} 
These quadratic terms give rise to the usual harmonic normal modes for a finite chain. In the case of equal masses $m_i = m$ the frequencies are

\begin{eqnarray} \label{HarmonicFreq}
\omega_{x, n} &=& 2 \sqrt{\frac{k_{\text{eff}}}{m}} \sin \left[ \frac{n \pi}{2 (N +1)} \right]  \nonumber \\ 
\omega_{y, n} &=& 2 \sqrt{\frac{k_{\perp}}{m}} \sin \left[ \frac{n \pi}{2 (N +1)} \right]
\end{eqnarray}
for $n=1$ to $N$. The frequencies in the $z$ direction are degenerated with those in $y$. In the case of no strain, the constant $k_{\perp}$ goes to zero, therefore, there are no harmonic modes in the transversal directions. The vibrations can be described by nonlinear modes whose frequencies are proportional to the amplitude and interact chaotically.

%%%%%%%%%%%%%%%%%%%%%%%%%%%%%%%%%%%%%%%%%%%%%%%%%%%%%%%%%%%%%%%%%%%%%%%%%
\section{Canonical calculation}
%%%%%%%%%%%%%%%%%%%%%%%%%%%%%%%%%%%%%%%%%%%%%%%%%%%%%%%%%%%%%%%%%%%%%%%%%

Considering that both thermal baths have the same temperature, the atomic chain is in thermal equilibrium, allowing a canonical calculation of the average displacement of the particles and their correlations. We start from the probability density of some possible state in the phase space:

\begin{equation}
{\cal P} (\{ \mathbf{r}_{i} \}, \{ \mathbf{p}_{i} \}) = \frac{1}{Z} \exp ( - \beta H (\{ \mathbf{r}_{i} \}, \{ \mathbf{p}_{i} \}) ) \ ,
\end{equation} 
with $\beta = 1/(\kB T)$, and the canonical partition function

\begin{equation} \label{PartitionFunction}
Z = {\cal C} \int \text{d}  \mathbf{p}_{1} \ldots  \text{d} \mathbf{p}_{N}  \text{d} \mathbf{r}_{1} \ldots  \text{d} \mathbf{r}_{N} \exp ( - \beta H (\{ \mathbf{r}_{i} \}, \{ \mathbf{p}_{i} \}) ) \ .
\end{equation} 
In the harmonic approximation, the Hamiltonian with potential energy (\ref{TotalPotApprox}) can be written as

\begin{equation}
H (\{ \mathbf{r}_{i} \}, \{ \mathbf{p}_{i} \} ) = \frac{1}{2} \sum_{i = 1}^{N} \frac{p_{x, i}^2 + p_{y, i}^2 + p_{z, i}^2 }{m_i} +  \frac{1}{2} \sum_{i = 1}^{N} \sum_{j = 1}^{N} [ k_{\text{eff}} x_i  \mathbb{K}_{i j} x_j + k_{\perp}  y_i  \mathbb{K}_{i j} y_j  + k_{\perp}  z_i  \mathbb{K}_{i j} z_j ] \ .
\end{equation}
The matrix $ \mathbb{K}$ is triadiagonal with $ \mathbb{K}_{i i} = 2,  \mathbb{K}_{i, i+1} = \mathbb{K}_{i+1, i} =  -1$, and zero otherwise. It can be easily diagonalized by the eigenvector unitary matrix

\begin{equation}
 \mathbb{A}_{k l} = \sqrt{\frac{2}{N+1}} \sin \left( \frac{\pi k l}{N+1} \right) \ ,
\end{equation}
with eigenvalues $\Omega_l = 2 \sin \left[ \frac{\pi l}{2(N+1)} \right]$. Changing to the eigenvector variables, all integrals in  (\ref{PartitionFunction}) are gaussian and can be performed. Finally the partition function is

\begin{equation} 
Z = \frac{ (2 \pi \kB T)^{3 N} }{ k_{\text{eff}}^{N/2} k_{\perp}^N  } \prod_{l = 1}^N \frac{m_l^{3/2}}{\Omega_l^3} \ .
\end{equation} 
With this result, and the same change of variables, we can compute the correlations, obtaining

\begin{equation} 
\langle x_i x_j \rangle = \frac{ \kB T }{ k_{\text{eff}}  } \sum_{l = 1}^N \frac{  \mathbb{A}_{i l}  \mathbb{A}_{j l} }{\Omega_l^2} =   \frac{ \kB T }{ k_{\text{eff}} }  \frac{ i (N+1 - j) }{ N + 1 }\ ,
\end{equation}
and similarly for

\begin{equation} 
\langle y_i y_j \rangle = \langle z_i z_j \rangle  =   \frac{ \kB T }{ k_{\perp} }  \frac{ i (N+1 - j) }{ N + 1 } \  .
\end{equation}
In this harmonic approximation of the potential, there are no correlations between different directions, and we can now compute the average distance between atoms. From (\ref{Distance}), up to second order we have

\begin{equation} 
\langle d_i  \rangle =  a + \langle  x_{i+1}-x_i   \rangle + \frac{1}{2 a} \langle \  (y_{i+1}-y_i)^2  + (z_{i+1}-z_i)^2  \rangle \  ,
\end{equation}
and replacing by the correlations we finally obtain

\begin{equation} \label{AverageD}
\langle d_i  \rangle =  a \left( 1  + \frac{N}{N+1}  \frac{ \kB T }{ k_{\perp} a^2 } \right) \  .
\end{equation}
The most important feature of this result is that the average distance between atoms does not depend on the bond index $i$, i.e. the thermal expansion is uniform along the nanowire. It also gives a leading term which is linear in temperature. This formula is for $  \kB T \ll k_{\perp} a^2$.

%%%%%%%%%%%%%%%%%%%%%%%%%%%%%%%%%%%%%%%%%%%%%%%%%%%%%%%%%%%%%%%%%%%%%%%%%
\section{Numerical results}
%%%%%%%%%%%%%%%%%%%%%%%%%%%%%%%%%%%%%%%%%%%%%%%%%%%%%%%%%%%%%%%%%%%%%%%%%

For higher temperatures, the displacements of the atoms are of the same order of the lattice constant, and the nonlinear terms become important. Also for the case of no tension, the potential in the transversal direction is nonlinear for all amplitudes. For these cases it is necessary to integrate the equations of motion numerically, taking into account the stochastic forces of the thermal baths. In all simulations we start from the equilibrium configuration, waiting for the system to attain a stationary regime before starting to perfom different statistics.

We consider the equilibrium distance $l_0$ between atoms as unit of length, the mass $m$ of atoms as unit of mass, $\tau_0 = \sqrt{m/k}$ as unit of time, $k l_0^2$ as unit of energy, and therefore $\Theta_0 = k l_0^2/\kB$  as unit of temperature. As an example, for carbon atoms in graphene $l_0 \approx$ 0.14 nm and $k \approx$ 650 nN/nm typically, giving $\Theta_0 \approx 10^6$ K and $1/\tau_0 \approx$ 180 THz for temperature and frequency units, respectively. Also expanding the Tersoff-Brenner potential around the equilibrium position up to fouth order, we obtain the dimensionless values $\alpha \approx -5.5$ and $\beta \approx 16.9$.

We study the thermal expansion of the system, by computing the temporal average distance between neighboring particles along the system. Although we perfom these calculations for different temperatures, strains, and strength of the cubic term of the potential, in all cases we find that the mean distance $\langle d_i  \rangle$ does not depend on the index $i$ significantly, beside some statistical errors, as it was shown in equation (\ref{AverageD}). Thus, the thermal expansion is uniform and particles near the thermal baths expand in the same way that particles in the middle of the nanowire. 

\begin{figure}[ht]
\begin{center}
\includegraphics[width=0.8\columnwidth]{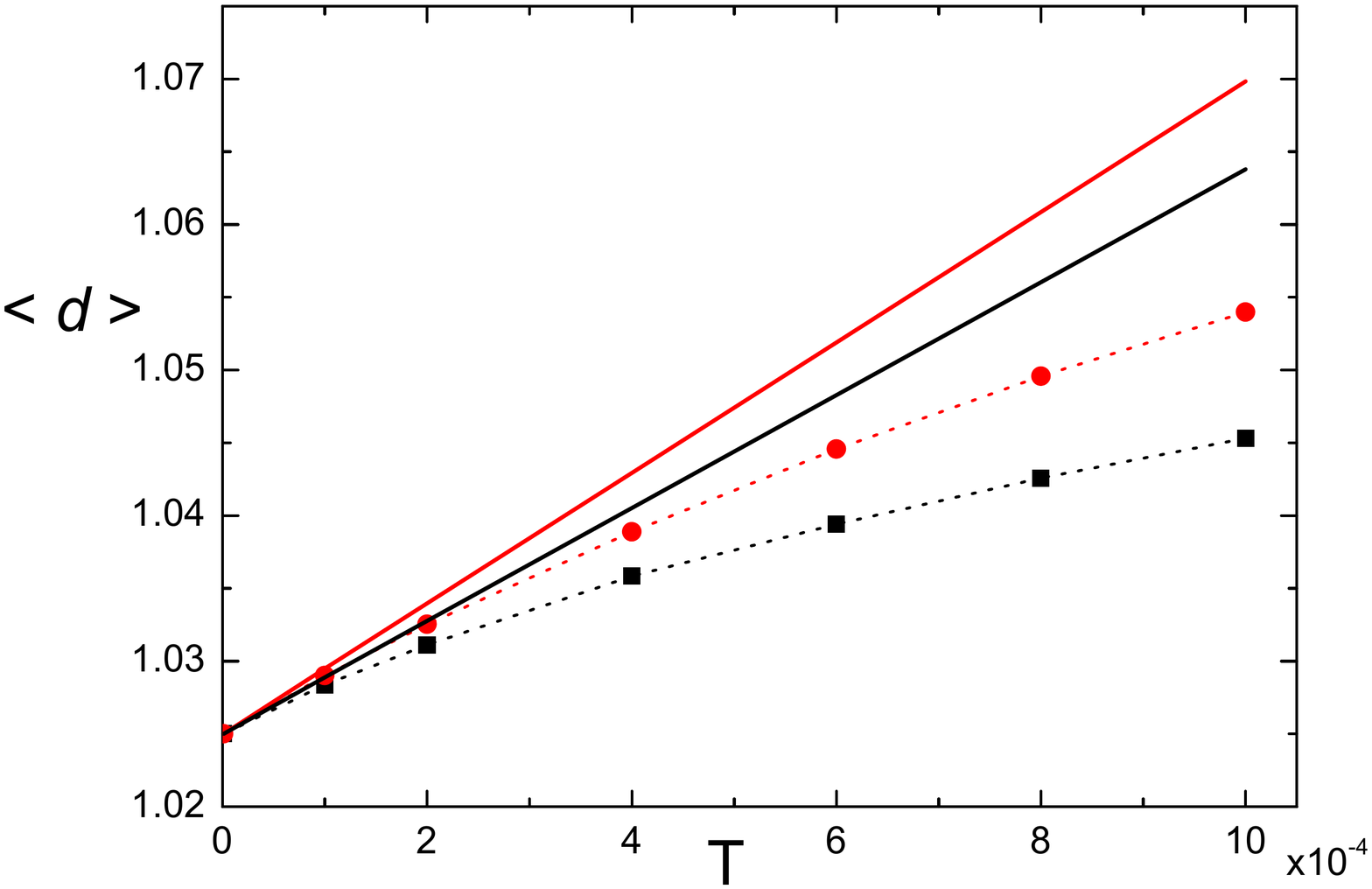}
\end{center}
\caption{Average distance between neighboring atoms $\langle d  \rangle$ as a function of temperature, for an homogeneous chain with $N = 50$, strain $\varepsilon$ = 0.025, and quartic term in the interatomic potential $\beta = 16.9$. Black squares and red circles are for cubic term $\alpha = 0$ and $\alpha = - 5.5$, respectively. Full lines are the canonical predictions given by equation (\ref{AverageD}), for the same two values of $\alpha$.}
\label{figDilatvsTemp}
\end{figure}

We analyze the temperature dependence of the thermal expansion in figure \ref{figDilatvsTemp}, for a fixed positive value of strain. As expected, we observe a general increasing of $\langle d \rangle$ for increasing temperature. For low temperatures, there is a linear relation as predicted by equation (\ref{AverageD}), the slope being the coefficient of thermal expansion. Nevertheless, at moderate and higher temperatures, the expansion has a decreasing slope, showing the effect of the nonlinear terms. We also plot the dependence for two different values of $\alpha$, the cubic term in the interatomic potential. A negative value of $\alpha$ implies a repulsive potential at short distances. Therefore it is not surprising that in this case we observe the biggest expansion in all temperature regimes, compared with $\alpha = 0$. We remark that even if the canonical calculation takes into account only harmonic terms in the potential, the dependence on $\alpha$ comes through $k_{\perp}$ after expanding the interatomic potential for a finite strain.

\begin{figure}[ht]
\begin{center}
\includegraphics[width=0.8\columnwidth]{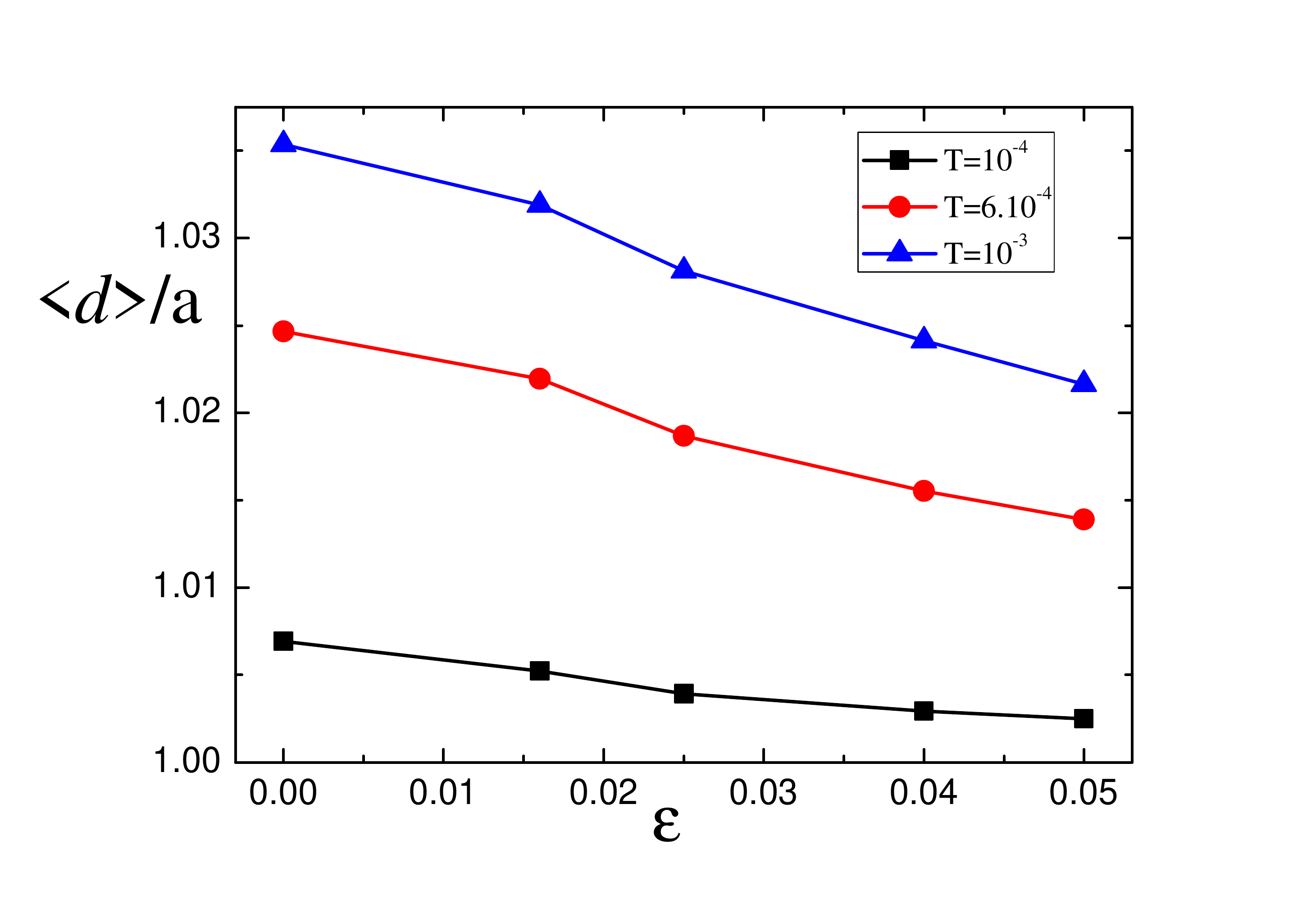}
\end{center}
\caption{Average distance between neighboring atoms $\langle d  \rangle$ as a function of strain, for an homogeneous chain with $N = 50$,  $\alpha= -5.5$, $\beta = 16.9$ and different temperatures. Black squares correspond to $T=10^{-4}$, red circles to $T=6.10^{-4}$ and blue triangles to $T=10^{-3}$}
\label{figDilatvsStrain}
\end{figure}

In Fig.\ref{figDilatvsStrain} we plot the thermal expansion as a function of strain, for three different temperatures. From now on we keep fixed $\alpha= -5.5$ (repulsive potential at short distances) and $\beta \approx 16.9$, which accounts for typical values between carbon atoms. Consistently with the previous figure, thermal expansion is bigger at higher temperature. But in all three regimes the average distance $\langle d \rangle$ decreases at increasing strain. This is an evidence that the chain becomes more rigid with tension, which is compatible with the increasing value of the transversal effective constant $k_{\perp}$.

In order to better understand the mechanical and thermal behavior, it is important to study the vibrational modes as a function of the different parameters. This would provide useful relations to use the system as a nanoresonator. 

\begin{figure}[ht]
\begin{center}
\includegraphics[width=0.8\columnwidth]{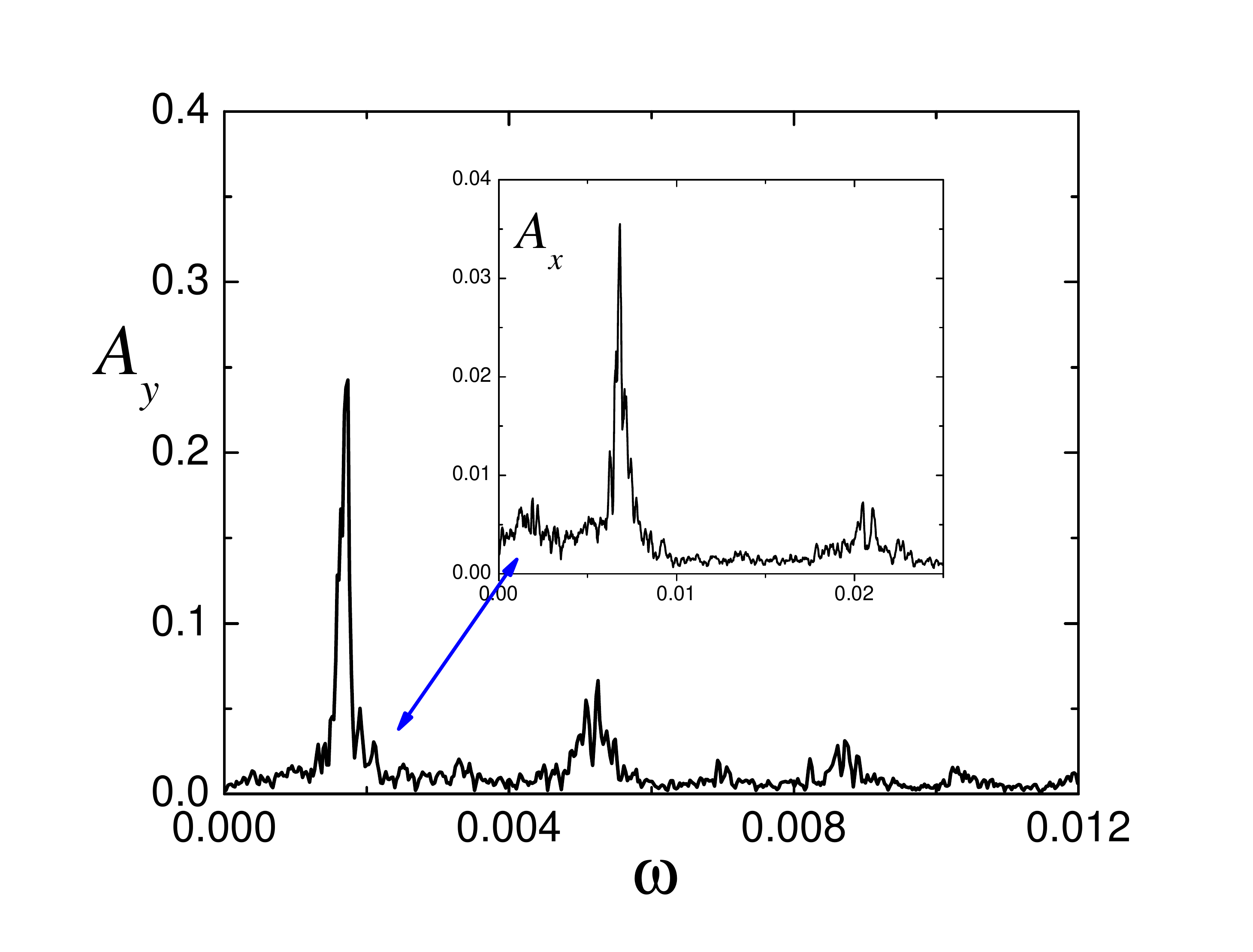}
\end{center}
\caption{Spectra of transversal and longitudinal (inset) displacements. $N = 50$, $\epsilon = 0.025$, $T = 6 \cdot 10^{-4}$}
\label{figSpectra}
\end{figure}

We perform Fourier transform of the longitudinal and transversal displacements of an atom in the chain, in order to find the main involved frequecies. In figure  \ref{figSpectra} we observe that most of the power is concentrated in the lowest transversal mode with $\omega_{\perp} \approx 0.0017$ (that would correspond to approximately 300 GHz for carbon atoms). The longitudinal lowest mode has a much higher frequency $\omega \approx 0.007$, but around an order of magnitude lower in power. It's also interesting to observe a small peak  around $\omega_{\perp}$ in the longitudinal spectrum. This means that both directions are effectively coupled by the nonlinear terms of the potential $c_3$ and $c_4$.

\begin{figure}[ht]
\begin{center}
\includegraphics[width=0.8\columnwidth]{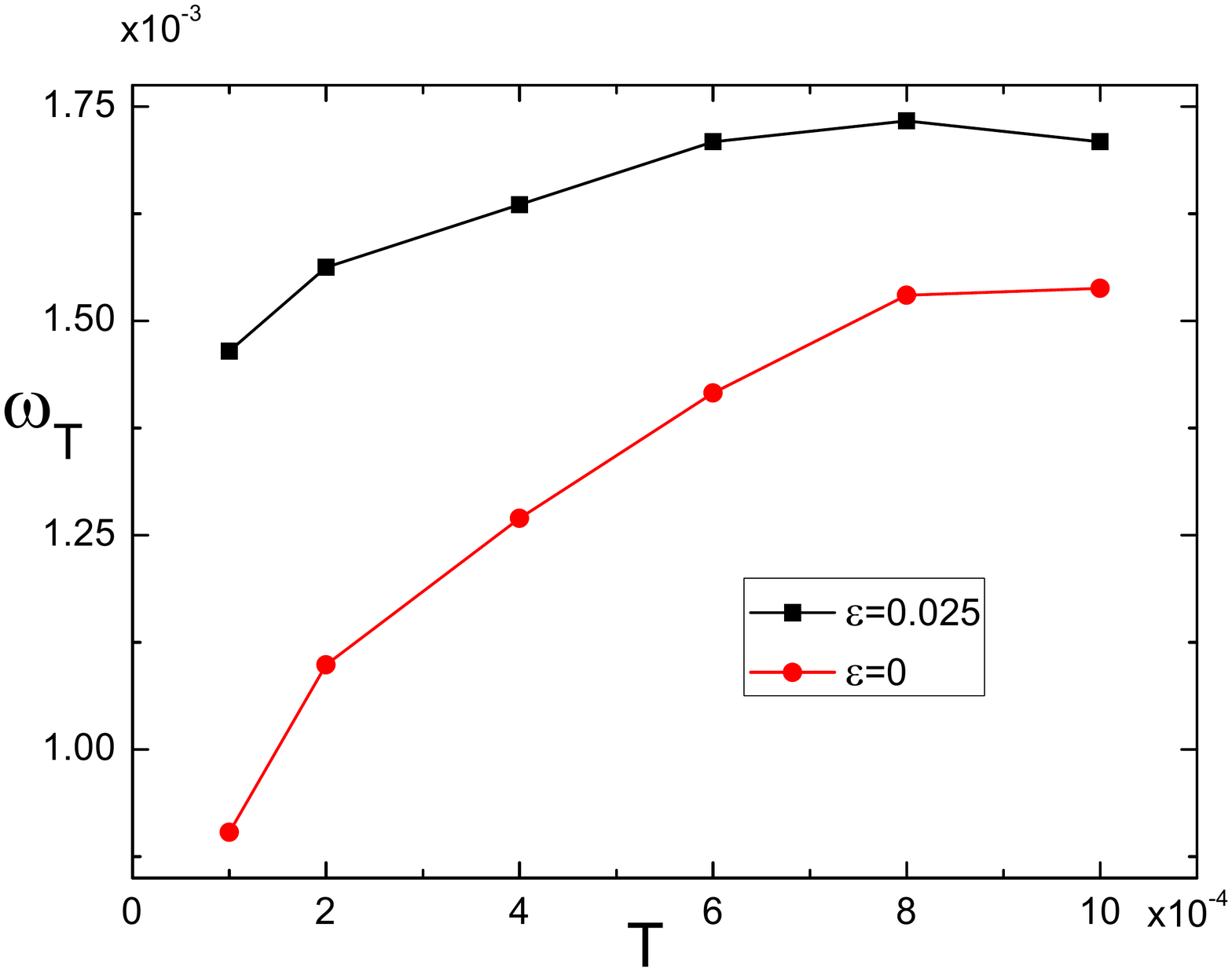}
\end{center}
\caption{Frequency of the lowest transversal mode as a function of temperature for $N = 50$ at two different strains.}
\label{figWvsTemp}
\end{figure}

In figure \ref{figWvsTemp} we observe an increase of $\omega_{\perp}$ with temperature. If harmonic normal modes of vibration are considered, their frequencies should not depend on available energy, i.e. the temperature. Therefore, the observed behavior is a consequence of nonlinear modes of vibration, whose frequencies depend on their amplitude. This effect is more pronounced when $\varepsilon = 0$, as $k_{\perp} = 0$, and the restitutive force in the transversal direction is proportional to the cube of the displacement.

\begin{figure}[ht]
\begin{center}
\includegraphics[width=0.8\columnwidth]{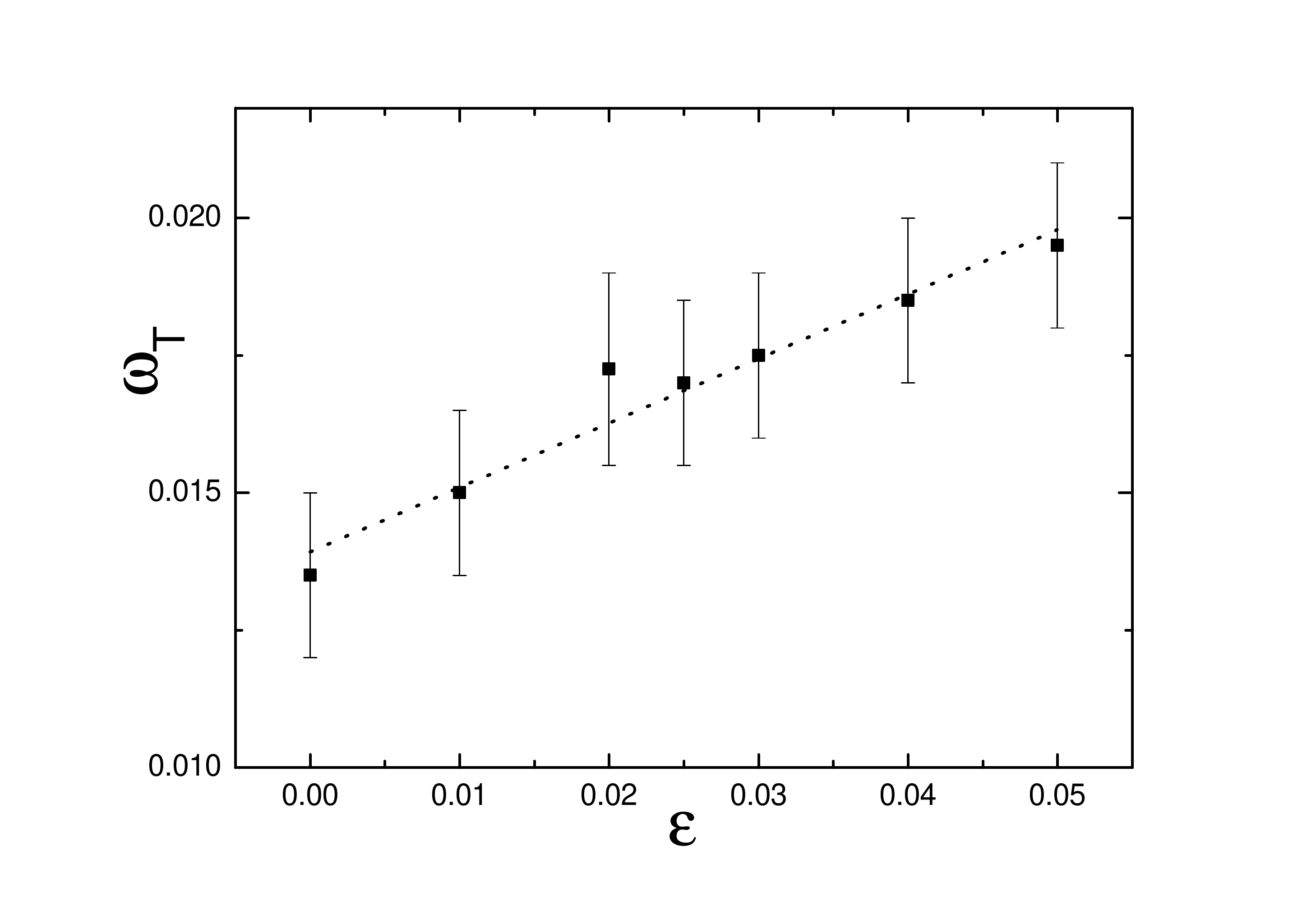}
\end{center}
\caption{Frequency of the lowest transversal mode as a function of strain, for $N = 50$ and $T = 6 \cdot 10^{-4}$ }
\label{figFreqvsStrain}
\end{figure}

We plot in figure \ref{figFreqvsStrain} the frequency $\omega_{\perp}$ as a function of strain, for a fixed temperature. We observe an approximate linear increasing. On one hand, this corresponds with the increasing value of $k_{\perp}$ with strain, that increases the rigidity and consequently the frequency of normal harmonic modes. On the other hand, at low strain the transversal modes are highly nonlinear, in which case their frequencies strongly depend on amplitude. From this linear behavior it is possible to calibrate the system as a strain sensor from measurements of the resonant frequency.

%%%%%%%%%%%%%%%%%%%%%%%%%%%%%%%%%%%%%%%%%%%%%%%%%%%%%%%%%%%%%%%%%%%%%%%%%
\section{Conclusions}
%%%%%%%%%%%%%%%%%%%%%%%%%%%%%%%%%%%%%%%%%%%%%%%%%%%%%%%%%%%%%%%%%%%%%%%%%

We have characterized thermal expansion and resonant frequencies in a nanowire at different temperature and strain regimes. We modeled the system as a one dimensional array, with interatomic potentials that depend on the absolute distance between atoms. However, the atoms can vibrate in the three directions, to consider a more general model. In turn, the interatomic potential besides a harmonic term includes nonlinear cubic and quartic terms (an $\alpha$-$\beta$ Fermi-Pasta-Ulam potential), as a general expansion of any potential.

Expanding the total potential energy around the equilibrium, effective coupling constants were obtained which depend on the strain  and can explain the interactions between the longitudinal and transversal vibrational modes. We performed a canonical calculation, obtaining a theoretical expresion for the thermal expansion, which is uniform along the nanowire and linear with temperature. There is also a dependence with strain through the effective transversal harmonic constant.

These theoretical results were compared to molecular dynamics simulations. For low temperatures up to $2 \cdot 10^{-4}$ (around 200 K for carbon atoms), the canonical calculation accurately describes the thermal expansion. For higher temperatures the expansion has a decreasing slope, pointing to a more relevant contribution of the nonlinear terms and a coupling between longitudinal and transversal directions. Nevertheless, it was checked that the thermal expansion is uniform along the system even at high temperatures and different strains.

We obtained the main resonant frequency of the system, that correspond to the lowest transversal vibrational mode, studying its dependence with temperature and strain. The dependence of the frequency with temperature shows that this vibrational mode is highly nonlinear. Moreover, we also found that it is linearly shifted when strain is applied, which allows to use this type of systems as nanoresonators. A challenging technological implication is the use of nanowires as sensors of nano-forces by inducing the system into a thermally nonlinear vibrational regime. Experimentally this can be achieved by measuring resonant frequencies from shifts observed in Raman spectroscopy \cite{raman2}.

More theoretical work should be done to better understand the nonlinear vibrational modes, the coupling between different modes, the localization phenomenon, and the implications in mechanical and thermal properties. The present work contributes in this direction and one of the authors (AM) will give a deeper insight to these aspects in a near future work.

In conclusion, the proposed atomistic model is a suitable approach to understand the underlying physics of nanosytems when transport properties are mediated by acoustic phonons. This model can also help to understand other thermal properties as conductance and thermal rectification phenomena.

\section*{Acknowledgments}
This work is supported by PIO-UNGS-CONICET grant.

\end{document}